# Spin-lasing in bimodal quantum dot micropillar cavities

*Niels Heermeier, Tobias Heuser, Jan Große, Natalie Jung, Arsenty Kaganskiy, Markus Lindemann, Nils C. Gerhardt, Martin R. Hofmann, Stephan Reitzenstein\**

N. Heermeier, T. Heuser, J. Große, Dr. A. Kaganskiy, Prof. S. Reitzenstein
Institut für Festkörperphysik
Technische Universität Berlin
Hardenbergstraße 36, 10623 Berlin, Germany
E-mail: stephan.reitzenstein@physik.tu-berlin.de

N. Jung, Dr. M. Lindemann, Prof. N. C. Gerhardt, Prof. M. R. Hofmann
Lehrstuhl für Photonik und Terahertztechnologie
Fakultät für Elektrotechnik und Informationstechnik
Ruhr-Universität Bochum, Universitätsstraße 150, 44780 Bochum, Germany

Spin-controlled lasers are highly interesting photonic devices and have been shown to provide ultra-fast polarization dynamics in excess of 200 GHz. In contrast to conventional semiconductor lasers their temporal properties are not limited by the intensity dynamics, but are governed primarily by the interaction of the spin dynamics with the birefringent mode splitting that determines the polarization oscillation frequency. Another class of modern semiconductor lasers are high-β emitters which benefit from enhanced light-matter interaction due to strong mode confinement in low-mode-volume microcavities. In such structures, the emission properties can be tailored by the resonator geometry to realize for instance bimodal emission behavior in slightly elliptical micropillar cavities. We utilize this attractive feature to demonstrate and explore spin-lasing effects in bimodal high-β quantum dot micropillar lasers. The studied microlasers with a β-factor of 4% show spin-laser effects with experimental polarization oscillation frequencies up to 15 GHz and predicted frequencies up to about 100 GHz which are controlled by the ellipticity of the resonator. Our results reveal appealing prospects for very compact, ultra-fast and energy-efficient spin-lasers and can pave the way for future purely electrically injected spin-lasers enabled by short injection path lengths.

1. Introduction

A high transmission bandwidth and stable transmission for optical communication systems are decisive for the Internet structure and the key to global digitization [1]. With Internet traffic and computing power increasingly concentrated in high-scale data centers due to the growing importance of cloud computing services, short-range optical communication systems play an important role [1]. These systems are mainly based on direct current modulated semiconductor lasers such as vertical-cavity surface-emitting lasers (VCSELs). However, the small-signal modulation bandwidth of VCSELs is limited to values below 40 GHz [2], [3], mainly due to the coupled carrier-photon dynamics in the resonator and additional parasitic electrical effects. It is highly questionable whether this technology will meet future bandwidth requirements. To close this gap, new concepts for ultra-fast short-range communication systems with higher modulation bandwidth are required. In this regard spin-lasers, such as spin-polarized VCSELs (spin-VCSELs), have recently proven to be a promising new device technology [4] [5] [6]. In these devices, the polarization state of the laser emission can be controlled by the spin state of the carriers and the transfer of angular momentum between carrier spin and photon spin plays a decisive role in their dynamical behavior [6]. In fact, the dynamics of the coupled spin system are usually decoupled from the intensity dynamics and can be much faster than those. Interestingly, the modulation bandwidth of spin-lasers can be directly controlled and increased by the mode splitting between the two orthogonal linearly polarized laser modes, e.g. by inducing birefringence into the resonator [7] [8] [9]. Frequencies > 200 GHz have been demonstrated following this concept recently which potentially provide a single channel data transmission rate of more than 240 Gbits/s [5]. Another important advantage of spin-VCSELs is their potential for very low power consumption. In contrast to conventional lasers, the ultra-fast modulation response in spin-lasers can be obtained for low carrier densities even close to threshold and is not severely affected by high temperatures [5]. This opens up new possibilities for energy-efficient high-speed communication systems, which are important for reducing the currently soaring energy consumption of data centers worldwide [1].

The need for novel energy-efficient devices can also be addressed by cavity-enhanced nano- and microlasers which have received significant scientific attention in recent years and which promise not only small size footprint but also strongly decreased threshold pump powers due to pronounced light-matter interaction [10]. In fact, in such devices pronounced light-matter interaction in the frame of cavity quantum electrodynamics (cQED) leads to a high fraction ($\beta$) of spontaneous emission coupled into the lasing mode, thus reducing the threshold pump powers by orders of magnitude in comparison to standard semiconductor lasers [11]. Moreover, high-β lasers are highly interesting from the fundamental science point of view because they allow one to explore the physical and technological limits of semiconductor lasers at the crossroads between classical and quantum physics [12]. Quantum dot (QD) micropillars are a very interesting type of high-β microlasers which often show an increased bimodal behavior due to a (usually unintentional) asymmetry of the pillar's cross section [13]. This leads to pronounced temporal mode switching and intriguing nonlinear dynamics effects [14], such as complex injection locking [15] and zero-lag synchronization [16], under external feedback or mutual coupling of micropillar cavities. Their bimodal behavior together with directional normal to the sample surface makes them highly interesting as a novel class of spin-lasers. In fact, the integration of ultra-fast spin and polarization dynamics with microcavities in bimodal QD-micropillar lasers is a novel concept combining the advantages of two emerging technologies. In particular using elliptically shaped bimodal micropillar cavities provides a perfect design parameter to induce the required mode splitting between orthogonally polarized laser modes, necessary to push the spin and polarization dynamics to highest frequencies.

In this work, we realize bimodal quantum dot micropillar cavities and we explore their potential to act as high-speed spin-lasers with well-controlled emission properties. We utilize the fact that the spectral splitting of the fundamental emission mode can be controlled by the ellipticity of the pillar's cross section which in turn controls the spin-oscillation frequency of these high-β microlasers. By implementing ellipticities $\varepsilon$ of up to 31 %, we realize mode splittings $\Delta E$ of up to 84 µeV (~21 GHz) to control the associated spin-oscillation frequencies. The experimental results are supported by theoretical modelling based on the spin-flip model [17] [6], which is a commonly used tool to predict spin-VCSEL behavior. Our work provides first insight into spin-lasing properties of high-β microlasers and has high potential to pave the way for highly compact, fast and energy efficient spin-lasers in the future.

2. Device fabrication and basic emission properties of bimodal QD-microlasers

2.1 Device fabrication

The bimodal micropillars under study are based on an AlGaAs/GaAs planar microcavity structure with a single layer of InGaAs QDs embedded in the central one-λ thick GaAs cavity. Using high-resolution electron beam lithography and plasma enhanced reactive ion etching we patterned arrays of micropillars with elliptical cross section (please see Methods section for details on the sample growth and device fabrication). To achieve the required bimodal emission behavior and to study the influence of the device geometry on the mode splitting we realized micropillars with ellipticities $\varepsilon = \sqrt{b/a} - 1$ up to 31%, where *a* (*b*) denotes the short (long) axis of the pillar's cross section. Figure 1 (a) shows a scanning electron microscope (SEM) image of three micropillars with a long axis of 4 µm and an ellipticity of about 10%. The structures show vertical sidewalls with rather small surface roughness. We etched about 1/3 of the lower distributed Bragg reflector (DBR) which is sufficient to achieve the desired lateral mode confinement and to maintain high quality (Q) factors in the range of 10.000 [18].

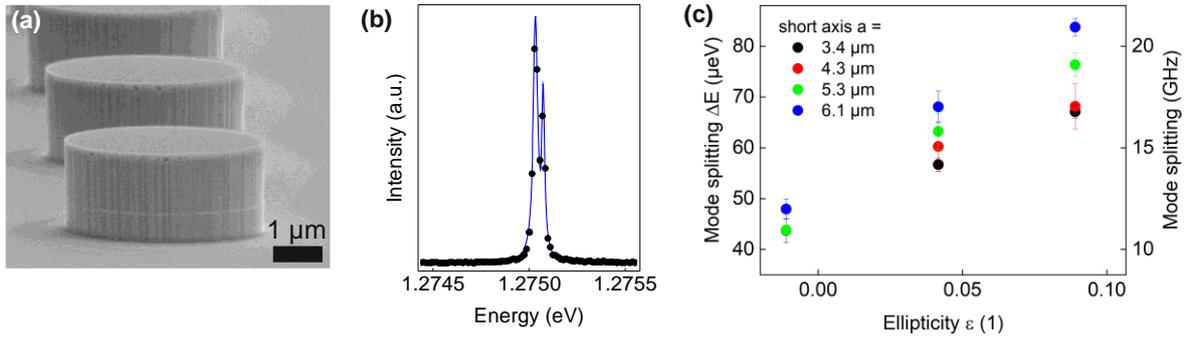

Figure 1: (a) SEM image of three elliptically shaped QD micropillar lasers with a long axis of 5.6 µm and an ellipticity of about 10%. (b) Corresponding µPL emission of an elliptical QD micropillar laser with a mode splitting of (41 ± 4) µeV and a diameter of 5.1 µm. (c) Dependence of the fundamental mode splitting in the ellipticity of the pillar cross section for different pillar diameters.

2.2 Basic emission properties

The emission properties of the QD micropillars were studied by high-resolution micro-photoluminescence (µPL) spectroscopy at low temperature (10K) (see Methods sections for details on the experimental setup). A µPL spectrum of a micropillar with a long axis of 5.6 µm and an ellipticity of about 10% is presented in Figure 1 (b). We observe two linearly polarized components of the fundamental emission mode with a spectral splitting *ΔE* of (41 ± 4) µeV induced by an elliptical cross-section. To obtain a better understanding on the influence of the intentionally introduced ellipticity we evaluated the resulting splitting *ΔE* of the fundamental mode in dependence of the pillar diameter and the ε parameter. Figure 1 (c) shows the measured mode splitting as function of the ellipticity *ε* for four families of pillars with different short axis *a* between 3.4 µm and 6.1 µm (determined by SEM investigations). For all values of *a* the desired mode splitting of the micropillar cavities increases almost linearly with the ellipticity *ε* up to values of approximately 80 µeV (~ 20 GHz) for the highest ellipticity of about 10% and a short axis of 6.1 µm. This dependence nicely shows that the mode splitting ΔE, and thus the expected spin-induced polarization oscillation frequency can be controlled precisely by the ellipticity of the pillar's cross section. Noteworthy, $\Delta E$ does not reach zero for a nominal $\varepsilon = 0$. This is explained by residual stress and strain, affecting the birefringence via the elasto-optic effect due to internal electrical fields [19].

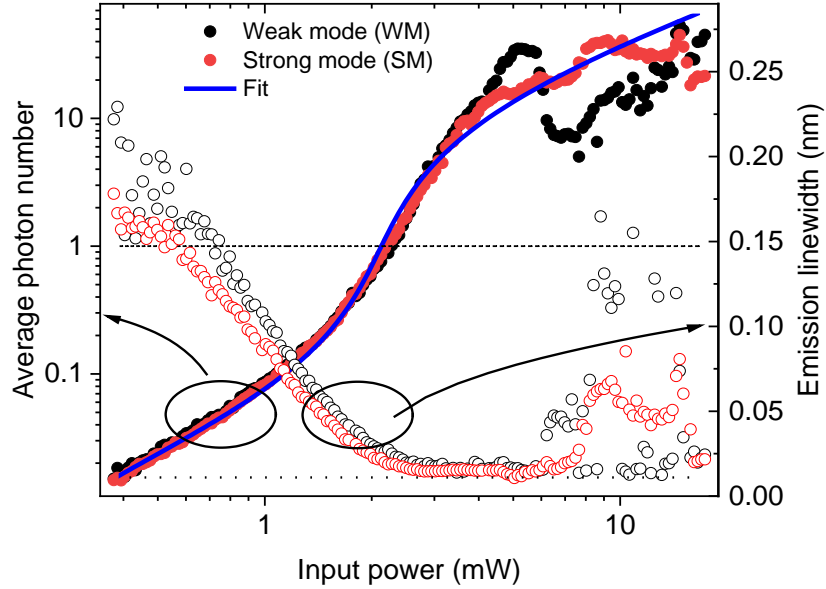

Figure 2: Input-output characteristics of a bimodal QD micropillar with a mode splitting of $(41 \pm 4)$ µeV and a diameter of 5.1 µm. The intensity is given in terms of average photon number $p$ which was obtained by fitting the experimental data. While the strong mode (SM) shows the typical s-shaped power dependence with some deviations in the high-power regime, the weak mode (WM) shows a drop in intensity at about 5 mW and a revival above 8 mW. The slight oscillatory behavior of mode intensity in the high excitation regime is explained in terms of gain competition of the two modes for the common QD gain. A fit (blue curve) to the data of the SW yields a β-factor of 4% and a threshold pump power of 2.1 mW, where the dashed horizontal line indicates $p = 1$ defined as the threshold photon number, see methods for details. The emission linewidth of both modes decreases strongly above 1 mW input power which confirms the onset of coherence at threshold, where the dotted horizontal line indicates the resolution limit of the spectrometer.

Next, we address the input-output dependence of the bimodal micropillar lasers under optical CW pumping at 785 nm. Figure 2 shows the emission intensity, in terms of the average photon number $p$ (the emission intensity is proportional to $p$), of the two fundamental mode components of the micropillar discussed above as function of the optical pump power in double logarithmic scale. The two emission modes feature very similar emission characteristics below threshold with an almost linear power scaling. At a pump power of about 2 mW the emission intensity starts to increase nonlinearly with the applied pump power and one mode, the strong mode (SM), shows the typical s-shaped behavior expected for high-β microlasers up to two times the threshold power. Fitting the data by a rate-equation model yields a β-factor of 4% and a threshold pump power of 2.1 mW for this mode (blue curve in Fig. 2). In the high excitation regime above approximately 5 mW the strong mode has higher intensity and shows slight oscillations in the emission intensity. Here, the intensity of the weak mode (WM) oscillates stronger and in general falls below the signal of the SM. The observed behavior at high excitation powers is typical for bimodal micropillar lasers and is explained by gain competition [20]. The corresponding power dependence of the emission linewidth depicted also in Fig. 2 confirms the transition to lasing for the SM with a strong linewidth narrowing at threshold due to enhanced temporal coherence in the lasing regime. Interestingly, the intensity fluctuations in the high excitation regime are also reflected in the emission linewidths and lead to deviations from the resolution limit for input powers exceeding 6 mW.

3. Spin-lasing of bimodal quantum dot microlasers

3.1 Time resolved detection of polarized emission

A central aspect of this work is the demonstration of spin-lasing in high-β microlasers. For this purpose, we pumped the microlaser discussed above optically by a CW laser at 785 nm, i.e. non-resonantly above the GaAs bandgap, at an excitation power of ~15 mW, which is about 7 times the threshold of 2.1 mW. This leads to an excitation of spin unpolarized carriers. For the injection of spin-polarized carriers and to access the dynamical properties, we applied in addition pulsed (ps-pulses at 80 MHz repetition rate) circularly polarized laser light with adjustable average power in the range of up to 10 mW at the sample surface. The light of the pulsed laser with a wavelength of 905 nm is resonant with wetting layer states of the QDs. Please see Methods for details on the experimental setup.

Polarization resolved measurements of emission from the bimodal QD micropillar using a streak camera with 4 ps time resolution allowed us to determine the time dependent emission in left and right circular polarizations denoted as $\boldsymbol{S^+}$ and $\boldsymbol{S^-}$, respectively. Figure 3(a) depicts the time resolved polarized emission detected from the QD-micropillar discussed above with a diameter of 5.1 µm and a mode splitting of (41 ± 4) µeV. The microlaser was driven with 14.98 mW CW excitation (at 785 nm) and additional pulsed circularly polarized excitation with 2.5 mW (at 905 nm). Upon excitation with the laser pulse at zero delay, both $\boldsymbol{S^+}$ (red trace) and $\boldsymbol{S^-}$ (black trace) rise and show an oscillatory behavior with increasing time delay. The associated oscillations in the circular polarization modes can clearly be resolved for delays larger than 100 ps. They show the expected antiphase behavior and are damped with a time constant of (0.19 ± 0.01) ns as determined from the sinusoidal fitting of the S3 parameter discussed in the next section.

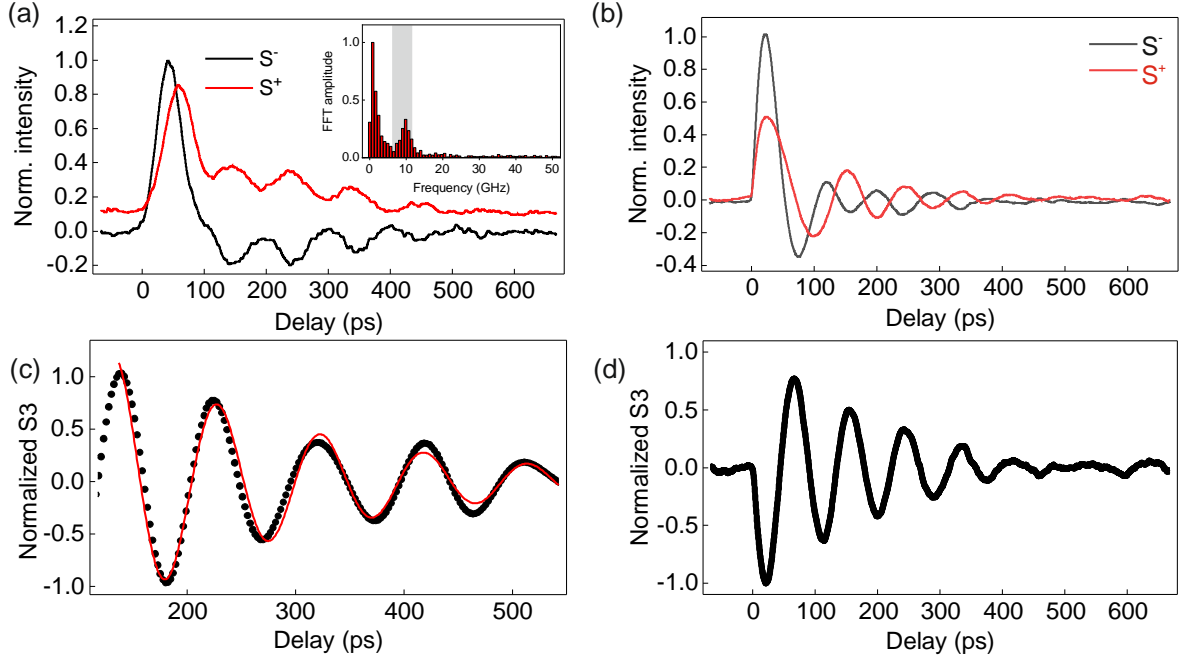

Figure 3: Spin polarized emission and spin-oscillation of a bimodal QD micropillar lasers with a mode splitting of (41 ± 4) µeV and a diameter of 5.1 µm recorded at pulsed laser power of 2.5 mW (CW power: 14.98 mW). (a) Time transient of the left (red) and right (black) circular polarized emission of the microlasers. (b) Simulation of the time transients of the left (red) and right (black) circular polarized emission of the microlaser utilizing the spin-flip model. CW offsets have been subtracted for presentation purposes. (c) S3 parameter calculated from the data presented in panel (a) after FFT analysis and applying a filter as indicated by the grey area in the inset. Fitting the data (dots) with a damped sinusoidal function (red, solid line) yields a polarization oscillation frequency of (10.4 ± 0.1) GHz in agreement with the corresponding FFT spectrum presented in the inset of panel (a). The decay constant of the S3 oscillation is (0.20 ± 0.01) ns. (d) S3 parameter calculated from the data presented in panel (b).

3.2 Time dependent spin-polarization degree

To obtain further insight into the spin-lasing properties of the QD-micropillar laser we determined the polarization degree of emission in terms of the S3 Stokes parameter, which is defined as $S3 = (S^+ - S^-)/(S^+ + S^-)$. The resulting time dependence of the S3 parameter after FFT filtering as indicated by the grey area in the inset of Fig. 3(a) is depicted in Fig. 3(c) for a pulsed laser power of 2.5 mW (black trace). The excitation with pulsed circularly polarized light induces an imbalance in the carrier spin-population for the spin-up and spin-down state. The observed oscillations reflect the resonance frequency of the coupled carrier-spin photon-spin system in the micropillar cavity. They can be observed in the S3 dynamics via angular momentum transfer between the carrier-spin and the photon-spin [6]. The oscillations can generally be understood as a beating between the two orthogonal linearly polarized fundamental modes in the micropillar cavity, frequency-splitted due to the micropillar anisotropies [5]. Fitting the experimental data after background subtraction with a damped sinusoidal function yields a spin-oscillation frequency of (10.4 ± 0.1) GHz and damping constant of (0.19 ± 0.01) ns. This frequency agrees with a local maximum at approximately 10 GHz in the FFT spectrum of the S3 time trace presented in the inset of Fig. 3(a). It is important to note that the extracted spin-oscillation frequency is in very good quantitative agreement with the micropillar mode splitting of (41 ± 4) μeV = (9.9 ± 1.0) GHz.

3.3 Comparison to the spin-flip model

We reproduce the measured spin-lasing utilizing the spin-flip model [17] (SFM), which was introduced to analyze the spin-lasing properties of VCSELs. In this work, we present simulations with a model parameterization adapted to micropillar lasers, i.e. with high-β, yielding very good agreement with experimental observations. Please see Methods for details on the simulation model and parameters.

The simulated traces of the left and right circular polarizations are depicted in Fig. 3(b). Upon excitation with the laser pulse at zero delay, both $S^+$ (red trace) and $S^-$ (black trace) rise and show an oscillatory behavior with increasing time delay, resembling the experimentally observed behavior with good agreement. The S3 Stokes parameter calculated for the simulated data is presented in Fig. 3(d). Like the measured oscillation, the oscillation frequency of S3 corresponds to the set micropillar mode splitting in the simulation.

3.4 Control of the polarization-oscillation frequency by mode splitting

An important advantage of spin-lasing in microlasers is the tight lateral mode confinement in these structures. As discussed above this mode confinement allows one to engineer the fundamental mode splitting of a bimodal micropillar which in turn is expected to control the polarization-oscillation frequency. In order to verify this prediction, we studied the spin-lasing properties for bimodal microlasers with ellipticity-induced mode splitting in the range from approximately 8.5 to 17.5 GHz. Figure 4 shows the resulting polarization-oscillation frequency $f_{PO}$ as function of the mode splitting $\Delta E$. We observe a clear increase of $f_{PO}$ with increasing $\Delta E$ which almost follows the expected linear dependence according to $f_{PO} = \Delta E$ as indicated by the dashed line. Furthermore, we simulate the dependence of the oscillation frequency on the mode splitting up to 80 GHz and find the same correspondence, as depicted in Fig. 4(b). Deviations from this direct correspondence can be explained with the dependence of the $f_{PO}$ on further laser properties such as photon density, photon lifetime, carrier recombination rate, spin-flip rate, linewidth enhancement factor and saturation effects, especially in the low birefringence regime used here [6].

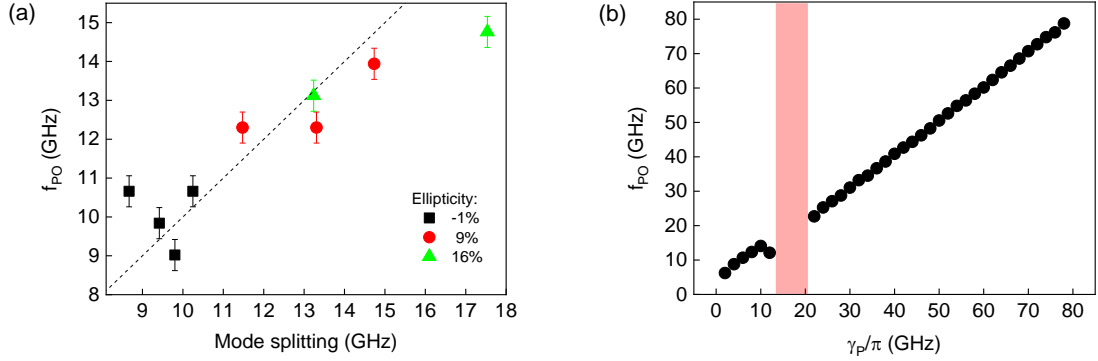

Figure 4: (a) Polarization-oscillation frequency $f_{PO}$ as function of the mode splitting $\Delta f$ of bimodal micropillars. A one-to-one correspondence $f_{PO} = \Delta f$ is indicated by the dashed line. (b) Spin-oscillation frequency $f_{PO}$ depending on the birefringence parameter used in the simulations. The range marked in red depicts a parameter combination, at which a more complicated and chaotic polarization behavior without damped polarization oscillations is observed.

Overall, the $f_{PO}(\Delta f)$ dependence highlights that the polarization-oscillation frequency can be predetermined by the fundamental mode splitting of micropillars which we engineer by the ellipticity of the pillar cross-section. As long as the dynamics of the coupled carrier-spin photon-spin system are sufficiently fast, the oscillations are expected to follow the mode splitting even to much higher frequencies. The simulations are based on the same set of parameters, except for the birefringence rate, that was used to describe the transient polarization dynamics in Fig. 3. The frequency range between 14-20 GHz, marked in red, reflects a region with different, more complex polarization dynamics including chaos. Here the simulated transients do not show damped polarization oscillations. Different dynamic regions in spin-VCSELs are known to depend strongly on device parameters such as spin flip rate, birefringence and dichroism [21]. For our set of parameters, the complex dynamics are suppressed for high birefringence rates where single period polarization oscillations can be observed again. The results for high birefringence rates show that even much higher oscillation frequencies of at least 80 GHz can be expected for the polarization dynamics of the micropillar lasers, indicating the potential of this novel type of high-ß spin-laser.

## 4. Conclusion

In conclusion we studied spin-lasing properties in high-β microlasers for the first time. Such lasers are highly interesting for the realization of the spin-lasing effect because of their compactness, low-power operation and the strong lateral mode confinement. The latter allowed us to induce a bimodal behavior in pillars with elliptical cross section. Here, the mode splitting of the fundamental mode is controlled by the ellipticity and reaches values up to 160 µeV (~40 GHz). By time resolved measurements of the bimodal micropillars we observed oscillatory behavior of circularly polarized emission and spin-oscillations with a frequency of up to 15 GHz. Here, the polarization-oscillation frequency can be controlled by the fundamental mode splitting which in turn is engineered by the elliptical geometry of the micropillar lasers. It is noteworthy, that mode splitting exceeding 1 meV (≈ 250 GHz) are achievable in strongly elliptical QD-micropillars [22] which gives great promises for ultra-fast high-β spin-microlasers. Overall, our results highlight the great potential of bimodal microlasers to act as spin-lasers with well controlled emission properties. Moreover, they will pave the way for further optimization of spin-lasing in ultra-small lasers with potentially unprecedented energy-efficiency and ultra-high spin-modulation frequency determined by the geometry of the microresonator.

## 5. Methods and Experimental Details

### 5.1. Sample Growth

The sample was grown via metalorganic vapor-phase epitaxy (MOVPE). The layers form a planar microcavity made up by a central one-λ thick GaAs cavity containing a single layer of InGaAs QDs sandwiched between a lower and an upper distributed Bragg reflector (DBR) composed of 27 (lower) and 23 (upper) λ/4-thick $Al_{0.9}Ga_{0.1}As$/GaAs mirror pairs. The QDs embedded in the middle of the central GaAs cavity are formed by self-assembled Stranski-Krastanow growth. To secure a high gain factor the QD growth is optimized for a sheet density of about $1\times10^{10}$ cm$^{-2}$ which is achieved through the deposition of 2.5 monolayers of $In_{0.6}Ga_{0.4}As$.

### 5.2 Micropillar Fabrication

The micropillar fabrication starts with the formation of a suitable etchmask material for the ICP-RIE etching process. To later realize steep and smooth sidewalls an etching recipe with a relative small etch rate of ~100 nm/min and a composition of $Cl_2$, $BCl_3$ and $Ar_2$ is chosen and therefore a dielectric SiN hardmask is needed, which provides the suitable selectivity for the etching process. At the first step the SiN gets deposited at an electrode temperature of 300°C using plasma enhanced chemical vapor deposition (PECVD) to form a 550 nm thick hardmask layer. Afterwards the sample is coated with a negative tone electron beam lithography resist (AZ nlof 2000 series, 0.5 µm Grade) of the same thickness. During the electron beam lithography process the sample gets exposed to the electron beam and the elliptical cross-section of the micropillars is written into the resist. After development, the resist mask is used to transfer the mask pattern into the hardmask layer by etching the SiN with a RIE process containing $SF_6$. With the defined hardmask the sample is now plasma etched with the ICP-RIE recipe. After etching the sample gets cleaned of the remains of the SiN by another RIE process using $CF_4$ as the etch gas.

5.3 Experimental Setup

The experimental setup is depicted in Figure 5. For the optical studies the sample is placed in a cryostat with moveable stage and a focusing lens in front of it. The excitation path is composed of two laser sources for optical pumping. A power adjustable diode laser which is driven in CW mode is used to operate the laser above threshold. Additionally, a mode-locked Titanium-sapphire (Ti:Sa) laser which is driven in ps pulsed mode is applied for optical spin injection. A logarithmic grey filter wheel at the output of the Ti:Sa laser allows us to tune the power fraction of pulsed optical excitation. The filter wheel is followed by a linear polarizer and a λ/4-plate which are adjusted to translate the linear polarization of the Ti:Sa laser to circular polarization, thus enabling optical spin injection. A beam splitter is used to combine the two excitation lasers to a common excitation path. A second grey filter wheel is added to the combined excitation path to tune the total optical pumping. This experimental configuration allows for a hybrid excitation scheme in which the generation of unpolarized and polarized charge carriers can conveniently be adjusted by the included filter elements.

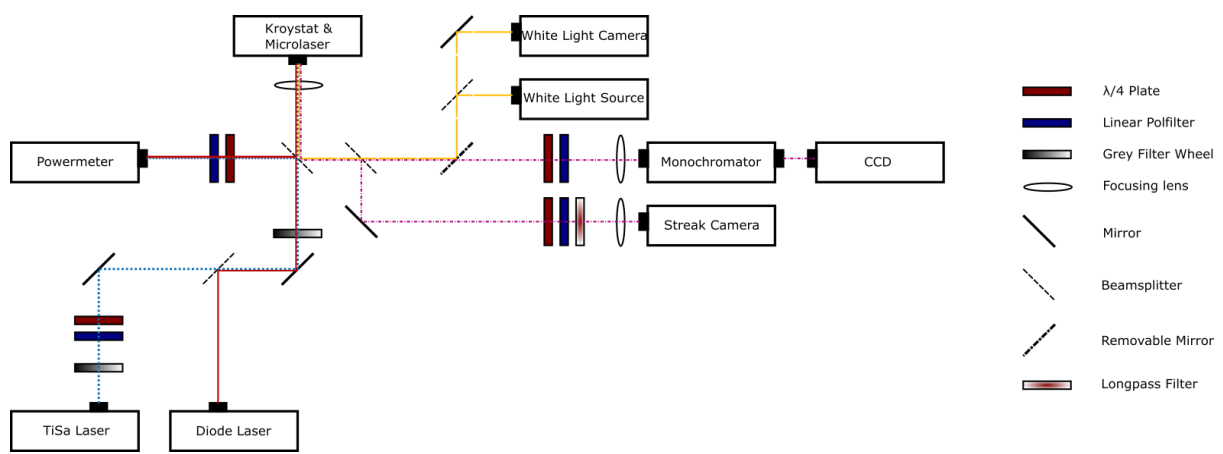

Figure 5: Schematic view of the experimental setup used to investigate spin-lasing effects of high-β microlasers.

Using a combination of beam splitters and mirrors three detection paths are created, each equipped with a λ/4-plate or a λ/2-plate and a linear polarizer for polarization selective analysis of microlaser emission. The first detection path diverts a (constant) fraction of the combined excitation path for power control. The second and third detection paths concern the signal of the sample, one of which is terminated with a monochromator with an attached CCD camera with a spectral resolution of 33 µeV while the other path is coupled to a streak camera with 4 ps time resolution. Hence, simultaneous time and spectrally resolved recording is possible in this configuration. An additional 950 nm long pass filter in front of the streak camera is used to isolate the micropillar signal from the spectral components of the optical pumping. Finally, a removable mirror can be added to the detection path to temporarily couple in a white light source and white light camera for adjusting the sample inside the cryostat.

**5.4 Rate-Equation Modelling of the Input-Output Dependence of the Microlaser**

A rate-equation approach is used for fitting the experimental data depicted in Fig. 2, and for extracting the mutually dependent microlaser parameters $\beta$ and $\zeta$. The model assumes negligible nonradiative recombination and relates the excitation intensity $I$ to the intracavity photon number $p$ as follows [23]:

$$I(p) = A * \left[\frac{p}{1+p} * (1+\zeta) * (1+\beta p) - \zeta \beta p\right],$$

which includes the parameters $\zeta = \frac{n_0 \beta}{\gamma \tau_{sp}}$, $\gamma = \frac{2\pi \nu_L}{Q}$ and $A = \frac{h\nu_L}{\tau_{ph} \beta \delta}$, with $n_0$ being the transparency carrier density, $\beta$ the spontaneous emission coupling factor, $\gamma$ the cavity decay rate, $\tau_{sp}$ and $\tau_{ph}$ the spontaneous emission and photon lifetime, respectively, $\nu_L$ the frequency of the lasing mode, $Q$ the cavity quality factor and $\delta$ the photon conversion efficiency [23], [24]. The fit function is found by solving the equation above for $p(I)$. A good fit of the experimental data can be achieved by $\zeta = 3$, $\beta = 4\%$ and $A = 0.5$ (see blue curve in Fig. 2). The fit reveals a threshold pumping power at $p = 1$ of $I_{th} = 2.1$ mW. Noteworthy, these values are in good agreement with those reported in [24] for very similar microlasers based on the same wafer material.

**5.5 Spin-Flip Model**

The spin-flip model is a set of coupled rate equations, which relate the electrical fields, the differential carrier density and the carrier-spin polarization utilizing several laser parameters [17] [25]. In accordance to the notation used in [25] we use the pumping term

$$\eta_\pm = \eta_0 \frac{1\pm P}{\sqrt{2\pi}} \exp\left(-\frac{2t^2}{\tau_p^2}\right) + \frac{J}{2}.$$

The list of parameters and their values can be found in Tab. 1. Gain compression was neglected and spontaneous emission noise was included using complex Gaussian white noise sources.

| Symbol | Parameter | Value | Dimension |
|---|---|---|---|
| $\alpha$ | Linewidth enhancement factor | 2 | |
| $\beta$ | β-Factor | 0.04 | |
| $\gamma$ | Carrier decay rate | 5 | ns$^{-1}$ |
| $\gamma_a$ | Linear dichroism | 3 | ns$^{-1}$ |
| $\gamma_p$ | Linear birefringence | 6.5π (Fig. 3) | GHz |
| | | 2π-78π (Fig. 4) | |
| $\gamma_s$ | Spin decay rate | 50 | ns$^{-1}$ |
| $\kappa$ | Photon decay rate | 50 | ns$^{-1}$ |
| J | Unpolarized bias pumping | 7 | (times threshold) |
| $\eta_0$ | Norm. pumping pulse amplitude | 12 | |
| P | Pumping polarization | -0.3 | |
| $\tau$ | Pumping pulse duration | 2.5×10$^{-3}$ | ns |

Acknowledgements

The research leading to these results has received funding the Volkswagen Foundation via the projects NeuroQNet2, from the European Research Council (ERC) under the European Union's Seventh Framework ERC Grant Agreement No. 615613 and from the German Research Foundation via projects 409799969, 250699912, and 392782903. We thank I. Žutić for valuable discussions.

Received: ((will be filled in by the editorial staff))
Revised: ((will be filled in by the editorial staff))
Published online: ((will be filled in by the editorial staff))